\documentclass[runningheads]{llncs}
\newcommand\blfootnote[1]{%
  \begingroup
  \renewcommand\thefootnote{}\footnote{#1}%
  \addtocounter{footnote}{-1}%
  \endgroup
}
\usepackage{graphicx}
\usepackage{amsmath}
\usepackage{adjustbox}
\usepackage{amssymb}
\usepackage{tabularx}
\usepackage{amsfonts}
\usepackage{amsmath}
\usepackage{hhline}
\usepackage{algorithmic}
\usepackage[ruled,vlined,linesnumbered]{algorithm2e}
\usepackage{derivative}
\usepackage{multirow}
\usepackage{float}
\usepackage{graphicx}
\usepackage{booktabs}
\usepackage{array}
\usepackage[misc,geometry]{ifsym}
\usepackage{etoolbox}

\usepackage[colorlinks=true,linkcolor=blue,urlcolor=blue,citecolor=blue]{hyperref}
\begin{document}
\title{DisC-Diff: Disentangled Conditional Diffusion Model for Multi-Contrast MRI Super-Resolution}

\author{Ye Mao\inst{1} \and
Lan Jiang\inst{2} \and 
Xi Chen \inst{3} \and 
Chao Li\inst{1,2,4}\textsuperscript{\Letter}} 
\titlerunning{DisC-Diff: Disentangled Conditional Diffusion Model for Super-Resolution}
\institute{Dept. of Clinical Neurosciences, University of Cambridge \\ \email{cl647@cam.ac.uk} \and
School of Science and Engineering, University of Dundee\and
Dept. of Computer Science, University of Bath\and
School of Medicine, University of Dundee 
\\\url{https://github.com/Yebulabula/DisC-Diff}}

\authorrunning{Y. Mao et al.}
\maketitle              
\begin{abstract}    
Multi-contrast magnetic resonance imaging (MRI) is the most common management tool used to characterize neurological disorders based on brain tissue contrasts. However, acquiring high-resolution MRI scans is time-consuming and infeasible under specific conditions. Hence, multi-contrast super-resolution methods have been developed to improve the quality of low-resolution contrasts by leveraging complementary information from multi-contrast MRI. Current deep learning-based super-resolution methods have limitations in estimating restoration uncertainty and avoiding mode collapse. Although the diffusion model has emerged as a promising approach for image enhancement, capturing complex interactions between multiple conditions introduced by multi-contrast MRI super-resolution remains a challenge for clinical applications. In this paper, we propose a disentangled conditional diffusion model, DisC-Diff, for multi-contrast brain MRI super-resolution. It utilizes the sampling-based generation and simple objective function of diffusion models to estimate uncertainty in restorations effectively and ensure a stable optimization process. Moreover, DisC-Diff leverages a disentangled multi-stream network to fully exploit complementary information from multi-contrast MRI, improving model interpretation under multiple conditions of multi-contrast inputs. We validated the effectiveness of DisC-Diff on two datasets: the IXI dataset, which contains 578 normal brains, and a clinical dataset with 316 pathological brains. Our experimental results demonstrate that DisC-Diff outperforms other state-of-the-art methods both quantitatively and visually.
\end{abstract}

\blfootnote{Y. Mao and L. Jiang contribute equally in this work.}

\section{Introduction}
Magnetic Resonance Imaging (MRI) is the primary management tool for brain disorders \cite{wei2022multi,wei2021structural,wei2021quantifying}. However, high-resolution (HR) MRI with sufficient tissue contrast is not always available in practice due to long acquisition time \cite{shi2015lrtv}, where low-resolution (LR) MRIs significantly challenge clinical practice.

Super-resolution (SR) techniques promise to enhance the spatial resolution of LR-MRI and restore tissue contrast. Traditional SR methods, e.g., bicubic interpolation \cite{khaledyan2020low}, compression sensing \cite{sen2009compressive}, and sparse representation \cite{yang2010image}, are based on non-learning techniques, which cannot restore the high-frequency details of images, i.e., textures, due to the ineffectiveness of establishing the complex non-linear mapping between HR and LR images. In contrast, deep learning (DL) has outperformed traditional methods, owing to its ability to capture fine details and preserve anatomical structures accurately.

Earlier DL-based SR methods \cite{chen2018brain,feng2021task,liu2018fusing,liu2019ganredl,shi2018super,wang2020enhanced,zhang2021mr} focused on learning the one-to-one mapping between the single-contrast LR MRI and its HR counterpart. However, multi-contrast MRI is often required for diagnosing brain disorders due to the complexity of brain anatomy. Single-contrast methods are limited by their ability to leverage complementary information from multiple MRI contrasts, leading to inferior SR quality. As an improvement, multi-contrast SR methods \cite{feng2021multi,lyu2020multi,stimpel2019multi,tsiligianni2021interpretable,zeng2018simultaneous} are proposed to improve the restoration of anatomical details by integrating additional contrast information. For instance, 
Zeng \textit{et al.} designed a CNN consisting of two subnetworks to achieve multi-contrast SR \cite{zeng2018simultaneous}.
Lyu \textit{et al.} presented a progressive network to generate realistic HR images from multi-contrast MRIs by minimizing a composite loss of mean-squared-error, adversarial loss, perceptual loss etc \cite{lyu2020multi}. Feng \textit{et al.} introduced a multi-stage integration network to extract complex interactions among multi-contrast features hierarchically, enhancing multi-contrast feature fusion \cite{feng2021multi}. Despite these advancements, most multi-contrast methods fail to 1) estimate restoration uncertainty for a robust model; 2) reduce the risk of mode collapse when applying adversarial loss to improve image fidelity.

Conditional diffusion models are a class of deep generative models that have achieved competitive performance in natural image SR \cite{dhariwal2021diffusion,li2022srdiff,rombach2022high}. The model incorporates a Markov chain-based diffusion process along with conditional variables, i.e., LR images, to restore HR images. The stochastic nature of the diffusion model enables the generation of multiple HR images through sampling, enabling inherent uncertainty estimation of super-resolved outputs. Additionally, the objective function of diffusion models is a variant of the variational lower bound that yields stable optimization processes. Given these advantages, conditional diffusion models promise to update MRI SR methods.

However, current diffusion-based SR methods are mainly single-contrast models. Several challenges remain for developing multi-contrast methods: 1) Integrating multi-contrast MRI into diffusion models increases the number of conditions. Traditional methods integrate multiple conditions 
via concatenation, which may not effectively leverage complementary information in multiple MRI contrasts, resulting in high-redundancy features for SR; 2) The noise and outliers in MRI can compromise the performance of standard diffusion models that use Mean Squared Error (MSE) loss to estimate the variational lower bound, leading to suboptimal results; 3) Diffusion models are often large-scale, and so are primarily intended for the generation of 2D images, i.e., treating MRI slices separately. Varied anatomical complexity across MRI slices can result in inconsistent diffusion processes, posing a challenge to efficient learning of SR-relevant features. 

To address the challenges, we propose a novel conditional disentangled diffusion model (DisC-Diff). To the best of our knowledge, this is the first diffusion-based multi-contrast SR method. The main contribution of our work is fourfold: 
\begin{itemize}
    \item We propose a new backbone network disentangled U-Net for the conditional diffusion model, a U-shape multi-stream network composed of multiple encoders enhanced by disentangled representation learning.
    \item We present a disentanglement loss function along with a channel attention-based feature fusion module to learn effective and relevant shared and independent representations across MRI contrasts for reconstructing SR images.
    \item We tailor a Charbonnier loss \cite{charbonnier1994two} to overcome the drawbacks of the MSE loss in optimizing the variational lower bound, which could provide a smoother and more robust optimization process.
    \item For the first time, we introduce an entropy-inspired curriculum learning strategy for training diffusion models, which significantly reduces the impact of varied anatomical complexity on model convergence.
\end{itemize}


Our extensive experiments on the IXI and in-house clinical datasets demonstrate that our method outperforms other state-of-the-art methods.

\section{Methodology}
\subsection{Overall Architecture}
\begin{figure}[ht]
\includegraphics[width=1.0\textwidth]{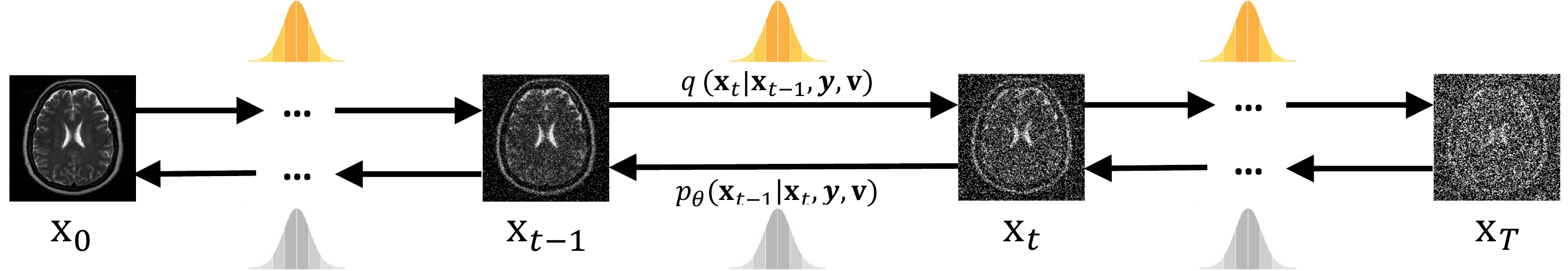}
\centering
\caption{Conceptual workflow of DisC-Diff on multi-contrast super-resolution. The forward diffusion process $q$ (left-to-right) perturbs HR MRI $\mathbf{x}$ by gradually adding Gaussian noise. The backward diffusion process $p$ (right-to-left) denoises the perturbed MRI, conditioning on its corresponding LR version $\mathbf{y}$ and other available MRI contrasts $\mathbf{v}$.}
\label{Fig1}
\end{figure}

The proposed DisC-Diff is designed based on a conditional diffusion model implemented in \cite{dhariwal2021diffusion}. As illustrated in Fig. \ref{Fig1}, the method achieves multi-contrast MRI SR through forward and reverse diffusion processes. Given an HR image $\mathbf{x}_0 \sim q(\mathbf{x}_0)$, the forward process gradually adds Gaussian noise to $\mathbf{x}_0$ over $T$ diffusion steps according to a noise variance schedule $\beta_1, \ldots, \beta_T$. Specifically, each step of the forward diffusion process produces a noisier image $\mathbf{x}_t$ with distribution $q(\mathbf{x}_t \mid \mathbf{x}_{t-1})$, formulated as:
\begin{gather}
q\left(\mathbf{x}_{1: T} \mid \mathbf{x}_{0}\right) = \prod_{t=1}^T q\left(\boldsymbol{x}_t \mid \boldsymbol{x}_{t-1}\right),
    q\left(\mathbf{x}_t \mid \mathbf{x}_{t-1}\right) = \mathcal{N}\left(\mathbf{x}_t ;\sqrt{1-\beta_t} \mathbf{x}_{t-1}, \beta_t \mathbf{I}\right)
\end{gather}

\begin{figure}[h]
\includegraphics[width=1.0\textwidth]{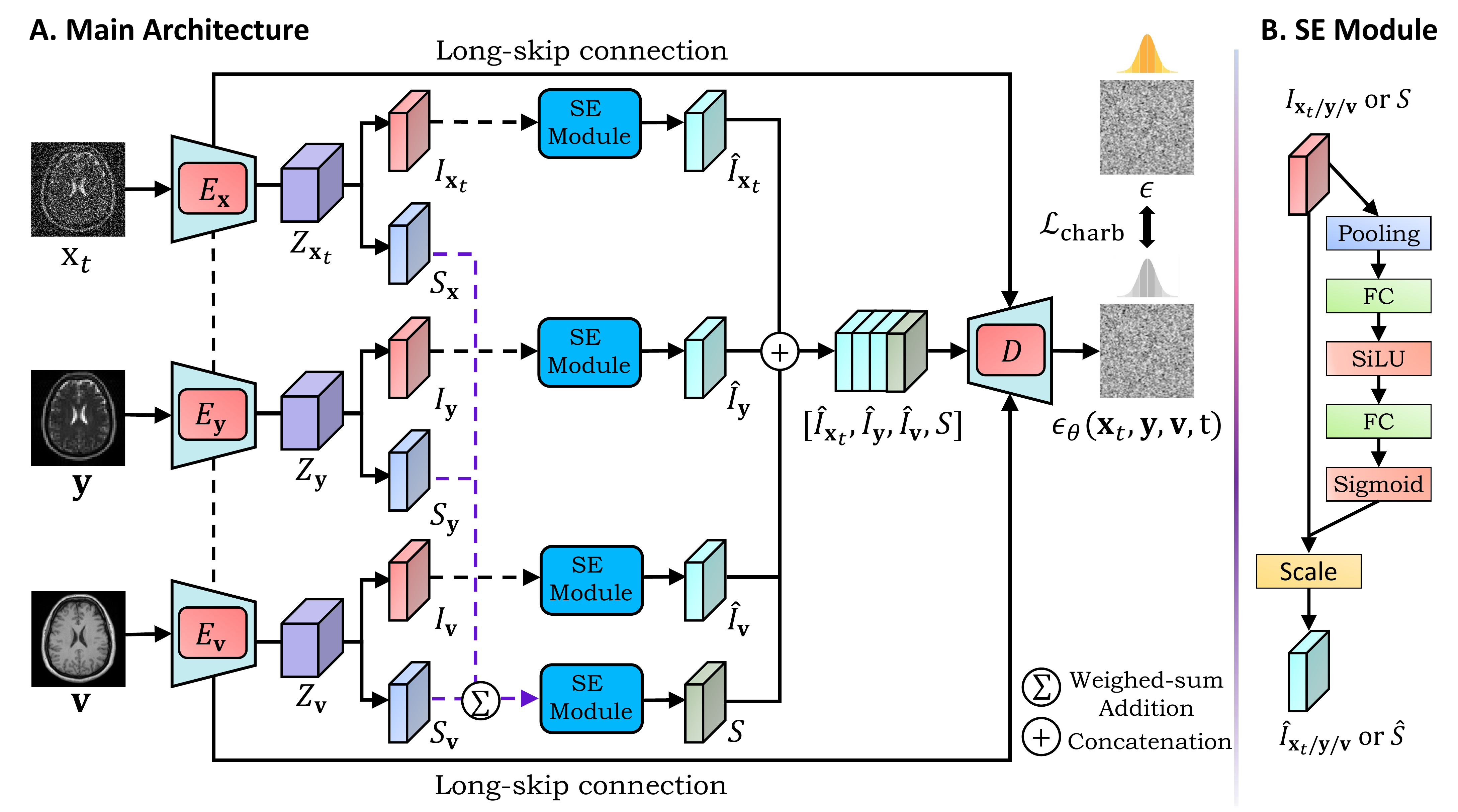}
\centering
\caption{(A) The disentangled U-Net consists of three encoders processing perturbed HR $\mathbf{x}_t$, LR $\mathbf{y}$, and additional contrast $\mathbf{v}$ respectively. The representations from each encoder are disentangled and concatenated as input to a single decoder $D$ to predict the intermediate noise level $\epsilon_\theta(\mathbf{x}_t, \mathbf{y}, \mathbf{v}, t)$. The architecture includes SE Modules detailed in (B) for dynamically weighting disentangled representations.} 
\label{Fig2}
\end{figure}

For sufficiently large $T$, the perturbed HR $\mathbf{x}_T$ can be considered a close approximation of isotropic Gaussian distribution. On the other hand, the reverse diffusion process $p$ aims to generate a new HR image from $\mathbf{x}_T$. This is achieved by constructing the reverse distribution $p_\theta\left(\mathbf{x}_{t-1}\mid\mathbf{x}_t, \mathbf{y}, \mathbf{v}\right)$, conditioned on its associated LR image $\mathbf{y}$ and MRI contrast $\mathbf{v}$, expressed as follows: 
\begin{gather}
p_\theta\left(\mathbf{x}_{0: T}\right)=p_\theta\left(\mathbf{x}_T\right) \prod_{t=1}^T p_\theta\left(\mathbf{x}_{t-1} \mid \mathbf{x}_t\right) \nonumber\\
p_\theta\left(\mathbf{x}_{t-1} \mid \mathbf{x}_t, \mathbf{y}, \mathbf{v} \right)=\mathcal{N}\left(\mathbf{x}_{t-1} ; \boldsymbol{\mu}_\theta\left(\mathbf{x}_t, \mathbf{y}, \mathbf{v}, t\right), \sigma^2_t\mathbf{I}\right)
\label{eq2}
\end{gather}
where $p_\theta$ denotes a parameterized model, $\theta$ is its trainable parameters and $\sigma^2_t$ can be either fixed to $\prod_{t=0}^t \beta_t$ or learned. It is challenging to obtain the reverse distribution via inference; thus, we introduce a disentangled U-Net parameterized model, shown in Fig. \ref{Fig2}, which estimates the reverse distribution by learning disentangled multi-contrast MRI representations. Specifically, $p_\theta$ learns to conditionally generate HR image by jointly optimizing the proposed disentanglement loss $\mathcal{L}_{\text{disent}}$ and a Charbonnier loss $\mathcal{L}_{\text{charb}}$. Additionally, we leverage a curriculum learning strategy to aid model convergence of learning
$\boldsymbol{\mu}_\theta\left(\mathbf{x}_t, \mathbf{y}, \mathbf{v}, t\right)$. 
\subsubsection{Disentangled U-Net} \label{sec1}
The proposed Disentangled U-Net is a multi-stream net composed of multiple encoders, separately extracting latent representations. 

We first denote the representation captured from the HR-MRI $\mathbf{x}_t$ as $Z_{\mathbf{x}_t} \in \mathbb{R}^{H \times W \times 2C}$, which contains a shared representation $S_{\mathbf{x}_t}$ and an independent representation $I_{\mathbf{x}_t}$ (both with 3D shape $H \times W \times C$) extracted by two $3\times3$ convolutional filters. The same operations on $\mathbf{y}$ and $\mathbf{v}$ yield $S_\mathbf{y}$, $I_\mathbf{y}$ and $S_\mathbf{v}$, $I_\mathbf{v}$, respectively. Effective disentanglement minimizes disparity among shared representations while maximizing that among independent representations. Therefore, $S_{{\mathbf{x}_t}/\mathbf{y}/\mathbf{v}}$ are as close to each other as possible and can be safely reduced to a single representation $S$ via a weighted sum, followed by the designed Squeeze-and-Excitation (SE) module (Fig. \ref{Fig2} B) that aims to emphasize the most relevant features by dynamically weighting the features in  $I_{{\mathbf{x}_t}/\mathbf{y}/\mathbf{v}}$ or $S$, resulting in rebalanced disentangled representations  $\hat{I}_{{\mathbf{x}_t}/\mathbf{y}/\mathbf{v}}$ and $\hat{S}$.
Each SE module applies global average pooling to each disentangled representation, producing a length-$C$ global descriptor. Two fully-connected layers activated by $\textbf{SiLU}$ and $\textbf{Sigmoid}$ are then applied to the descriptor to compute a set of weights $\mathbf{s} = [s_1, \ldots, s_i, s_{i+1}, \ldots, s_C]$, where $s_i$ represents the importance of the $i$-th feature map in the disentangled representation. 
Finally, the decoder block $D$ shown in Fig. \ref{Fig2} performs up-sampling on the concatenated representations $[\hat{I}_{\mathbf{x}_t}, \hat{I}_{\mathbf{y}}, \hat{I}_{\mathbf{v}}, \hat{S}]$ and outputs a noise prediction $\epsilon_\theta(\mathbf{x}_t, \mathbf{y}, \mathbf{v}, t)$ to compute the Gaussian mean in Eq. \ref{eq3}:
\begin{gather}
\boldsymbol{\mu}_\theta\left(\mathbf{x}_t,\mathbf{y}, \mathbf{v}, t\right)=\frac{1}{\sqrt{\alpha_t}}\left(\mathbf{x}_t-\frac{\beta_t}{\sqrt{1-\bar{\alpha}_t}} \boldsymbol{\epsilon}_\theta\left(\mathbf{x}_t, \mathbf{y}, \mathbf{v}, t\right)\right)
\label{eq3}
\end{gather}
where $\alpha_t = 1-\beta_t$ and $\bar{\alpha}_t = \prod^t_{s=0}\alpha_s$.

\subsection{Design of loss functions}
To effectively learn disentangled representations with more steady convergence in model training, a novel joint loss is designed in DisC-Diff as follows. 
\subsubsection{Disentanglement Loss} $\mathcal{L}_\text{disent}$ is defined as a ratio between $\mathcal{L}_\text{shared}$ and $\mathcal{L}_\text{indep}$, where $\mathcal{L}_\text{shared}$ measures the $\mathcal{L}_2$ distance between shared representations, and $\mathcal{L}_\text{indep}$ is the distance between independent representations:

\begin{align}
\mathcal{L}_{\text {disent }} = \frac{\mathcal{L}_{\text {shared }}}{\mathcal{L}_{\text {indep}}}  
= \frac{\|S_{\mathbf{x}_t} - S_\mathbf{y}\|_2 + \|S_{\mathbf{x}_t} - S_\mathbf{v}\|_2 + \|S_\mathbf{y} - S_\mathbf{v}\|_2}{\|I_{\mathbf{x}_t} - I_\mathbf{y}\|_2 + \|I_{\mathbf{x}_t} - I_\mathbf{v}\|_2 + \|I_\mathbf{y} - I_\mathbf{v}\|_2}
\end{align}


\subsubsection{Charbonnier Loss} $\mathcal{L}_\text{charb}$ is a smoother transition between $\mathcal{L}_1$ and $\mathcal{L}_2$ loss, facilitating more steady and accurate convergence during training \cite{lai2018fast}. It is less sensitive to the non-Gaussian noise in $\mathbf{x}_t$, $\mathbf{y}$ and $\mathbf{v}$, and encourages sparsity results, preserving sharp edges and details in MRIs. It is defined as:
\begin{equation}
    \mathcal{L}_\text{charb} = \sqrt{(\epsilon_\theta(\mathbf{x}_t, \mathbf{y}, \mathbf{v}, t) - \epsilon)^2 + \gamma^2)}
\end{equation}
where $\gamma$ is a known constant. The total loss is the weighted sum of the above two losses:
\begin{equation}
    \mathcal{L}_\text{total} = \lambda_1 \mathcal{L}_\text{disent} + \lambda_2 \mathcal{L}_\text{charb}
\end{equation}
where $\lambda_1$, $\lambda_2 \in (0,1]$ indicate the weights of the two losses.



\subsection{Curriculum Learning} Our curriculum learning strategy improves the disentangled U-Net's performance on MRI data with varying anatomical complexity by gradually increasing the difficulty of training images, facilitating efficient learning of relevant features. All MRI slices are initially ranked based on the complexity estimated by Shannon-entropy values of their ground-truth HR-MRI, denoted as an ordered set $E =\{e_\text{min}, \ldots, e_\text{max}\}$. Each iteration samples $N$ images whose entropies follow a normal distribution with $e_\text{min}<\mu<e_\text{max}$. As training progresses, $\mu$ gradually increases from $e_\text{min}$ to $e_\text{max}$, indicating increased complexity of the sampled images. The above strategy is used for the initial $M$ iterations, followed by uniform sampling of all slices.
 \section{Experiments \& Results}
\subsubsection{Datasets \& Baselines}
We evaluated our model on the public IXI dataset\footnote{\url{https://brain-development.org/ixi-dataset/}} and an in-house clinical brain MRI dataset. In both datasets, our setting is to utilize  HR T1-weighted images $\text{HR}_{T1}$ and LR T2-weighted image $\text{LR}_{T2}$ created by $k$-space truncation \cite{chen2018brain} to restore $2\times$ and $4\times$ HR T2-weighted images, aligning with the setting in \cite{feng2021multi,tsiligianni2021interpretable,zeng2018simultaneous}.
 
We split the 578 healthy brain MRIs in the IXI dataset into 500 for training, 6 for validation, and 70 for testing. We apply center cropping to convert each MRI into a new scan comprising 20 slices, each with a resolution 224 $\times$ 224. The processed IXI dataset is available for download at this link \footnote{\url{https://bit.ly/3yethO4}}.
The clinical dataset is fully sampled using a 3T Siemens Magnetom Skyra scanner on 316 glioma patients. The imaging protocol is as follows: $\text{TR}_{T1}$ = 2300 ms, $\text{TE}_{T1}$ = 2.98 ms, FOV$_{T1}$ = 256 $\times$ 240 mm$^2$, $\text{TR}_{T2}$ = 4840 ms, $\text{TE}_{T2}$ = 114 ms, and FOV$_{T2}$ = 220 $\times$ 165 mm$^2$. The clinical dataset is split patient-wise into train/validation/test sets with a ratio of 7:1:2, and each set is cropped into $\mathbb{R}^{224 \times 224 \times 30}$. 

We compare our method with three single-contrast SR methods (bicubic interpolation, EDSR \cite{lim2017enhanced}, SwinIR \cite{liang2021swinir}) and three multi-contrast SR methods (Guided Diffusion \cite{dhariwal2021diffusion}, MINet \cite{feng2021multi}, MASA-SR \cite{lu2021masa}).

\subsubsection{Implementation Details} \label{imp} DisC-Diff was implemented using PyTorch with the following hyperparameters: $\lambda_1 = \lambda_2 = 1.0$, diffusion steps $T$ = $1000$, 96 channels in the first layer, 2 BigGAN Residual blocks, and attention module at 28$\times$28, 14$\times$14, and 7$\times$7 resolutions. The model was trained for 200,000 iterations ($M$ = 20,000) on two NVIDIA RTX A5000 24 GB GPUs using the AdamW optimizer with a learning rate of $10^{-4}$ and a batch size of 8. Following the sampling strategy in \cite{dhariwal2021diffusion}, DisC-Diff learned the reverse diffusion process variances to generate HR-MRI in only 100 sampling steps. The baseline methods were retrained with their default hyperparameter settings. Guided Diffusion was modified to enable multi-contrast SR by concatenating multi-contrast MRI     as input.

\subsubsection{Quantitative Comparison}
The results show that DisC-Diff outperforms other evaluated methods on both datasets at 2$\times$ and 4$\times$ enlargement scales. Specifically, on the IXI dataset with 4$\times$ scale, DisC-Diff achieves a PSNR increment of 1.44 dB and 0.82 dB and an SSIM increment of 0.0191 and 0.0134 compared to state-of-the-art single-contrast and multi-contrast SR methods \cite{liang2021swinir,lu2021masa}. The results show that without using disentangled U-Net as the backbone, Guided Diffusion performs much poorer than MINet and MASA-SR on the clinical dataset, indicating its limitation in recovering anatomical details of pathology-bearing brain. Our results suggest that disentangling multiple conditional contrasts could help DisC-Diff accurately control the HR image sampling process. Furthermore, the results indicate that integrating multi-contrast information inappropriately may damage the quality of super-resolved images, as evidenced by multi-contrast methods occasionally performs worse than single-contrast methods, e.g., EDSR showing higher SSIM than MINet on both datasets at 2$\times$ enlargement. 
\begin{table*}[t]
\centering
\caption{Quantitative results on both datasets with $2\times$ and $4\times$ enlargement scales in terms of mean PSNR (dB) and SSIM. Bold numbers indicate the best results.}
\begin{adjustbox}{max width=0.9\textwidth}
\begin{tabular}{c||cccc|cccc}
\toprule
\hline
Dataset          & \multicolumn{4}{c|}{IXI}                                                                 & \multicolumn{4}{c}{Clinical Dataset}                                                    \\ \hline
Scale            & \multicolumn{2}{c|}{$2\times$}                               & \multicolumn{2}{c|}{$4\times$}          & \multicolumn{2}{c|}{$2\times$}                               & \multicolumn{2}{c}{$4\times$}          \\ \hline 
Metrics          & \multicolumn{1}{c}{PSNR} & \multicolumn{1}{c|}{SSIM} & \multicolumn{1}{c}{PSNR} & SSIM & \multicolumn{1}{c}{PSNR} & \multicolumn{1}{c|}{SSIM} & \multicolumn{1}{c}{PSNR} & SSIM \\ \hline \hline
Bicubic          & \multicolumn{1}{c}{32.84}     & \multicolumn{1}{c|}{0.9622}     & \multicolumn{1}{c}{26.21}     & 0.8500     & \multicolumn{1}{c}{34.72}     & \multicolumn{1}{c|}{0.9769}     & \multicolumn{1}{c}{27.17}     & 0.8853     \\ 
EDSR \cite{lim2017enhanced}             & \multicolumn{1}{c}{36.59}     & \multicolumn{1}{c|}{0.9865}     & \multicolumn{1}{c}{29.67}     & 0.9350     & \multicolumn{1}{c}{36.89}     & \multicolumn{1}{c|}{0.9880}     & \multicolumn{1}{c}{29.99}     & 0.9373     \\ 
SwinIR \cite{liang2021swinir}             & \multicolumn{1}{c}{37.21}     & \multicolumn{1}{c|}{0.9856}     & \multicolumn{1}{c}{29.99}     & 0.9360     & \multicolumn{1}{c}{37.36}     & \multicolumn{1}{c|}{0.9868}     & \multicolumn{1}{c}{30.23}     & 0.9394      \\
Guided Diffusion \cite{dhariwal2021diffusion} & \multicolumn{1}{c}{36.32}     & \multicolumn{1}{c|}{0.9815}     & \multicolumn{1}{c}{30.21}     & 0.9512     & \multicolumn{1}{c}{36.91}     & \multicolumn{1}{c|}{0.9802}     & \multicolumn{1}{c}{29.35}     & 0.9326     \\ 
MINet  \cite{feng2021multi}          & \multicolumn{1}{c}{36.56}     & \multicolumn{1}{c|}{0.9806}     & \multicolumn{1}{c}{30.59}     & 0.9403     & \multicolumn{1}{c}{37.73}     & \multicolumn{1}{c|}{0.9869}     & \multicolumn{1}{c}{31.65}     &  0.9536    \\ 
MASA-SR \cite{lu2021masa}         & \multicolumn{1}{c}{-}     & \multicolumn{1}{c|}{-}     & \multicolumn{1}{c}{30.61}     & 0.9417      & \multicolumn{1}{c}{-}     & \multicolumn{1}{c|}{-}     & \multicolumn{1}{c}{31.56}     &  0.9532    \\
\hline \hline
\textbf{DisC-Diff (Ours)}  & \multicolumn{1}{c}{\textbf{37.64}}     & \multicolumn{1}{c|}{\textbf{0.9873}}     & \multicolumn{1}{c}{\textbf{31.43}}     & \textbf{0.9551}     & \multicolumn{1}{c}{\textbf{37.77}}     & \multicolumn{1}{c|}{\textbf{0.9887}}     & \multicolumn{1}{c}{\textbf{32.05}}     & \textbf{0.9562}    \\ \hline
\bottomrule
\end{tabular}
\end{adjustbox}
\label{tab1}
\end{table*}

\begin{figure}[h]
    \centering
\includegraphics[width=1.0\textwidth]{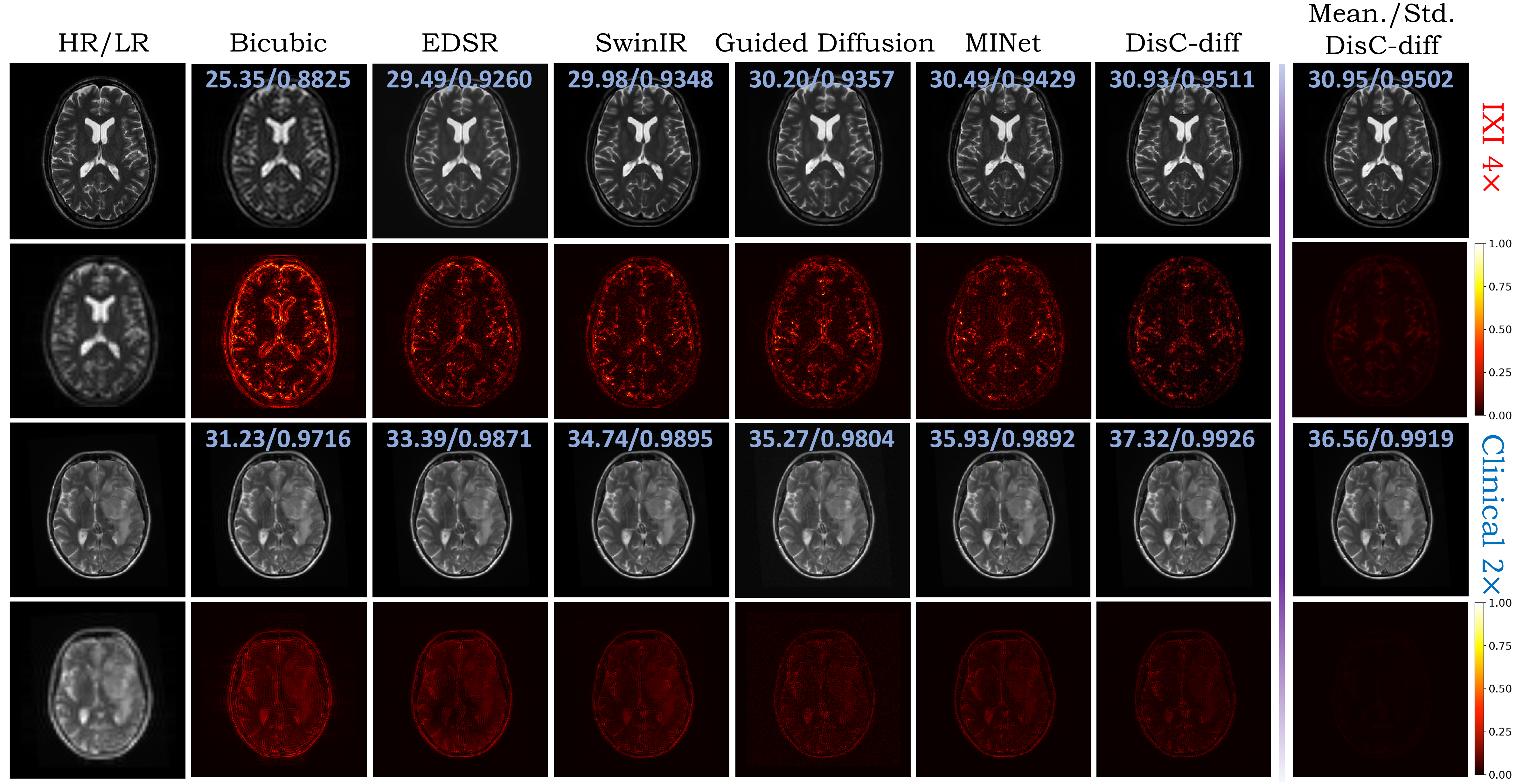}
    \caption{Visual restoration results and error maps of different methods on IXI ($4\times$) and glioma ($2\times$) datasets, along with mean and standard deviation of our method's sampling results for indicating super-resolution uncertainty (Last column). }
    \label{fig:3}
\end{figure}
\subsubsection{Visual Comparison and Uncertainty Estimation}
Fig. \ref{fig:3} shows the results and error maps for each method under IXI (2$\times$) and Clinical (4$\times$) settings, where less visible texture in the error map indicates better restoration. DisC-Diff outperforms all other methods, producing HR images with sharper edges and finer details, while exhibiting the least visible texture. Multi-contrast SR methods consistently generate higher-quality SR images than single-contrast SR methods, consistent with their higher PSNR and SSIM. Also, the lower variation between different restorations at the 2$\times$ scale compared to the 4$\times$ scale (Last column in Fig. \ref{fig:3}) suggests higher confidence in the 2$\times$ restoration results.

\subsubsection{Ablation Study} We assess the contribution of three key components in DisC-Diff: 1) $w/o$ $\mathcal{L}_{\text{disent}}$ - implement our model without disentanglement loss, 2) $w/o$ $\mathcal{L}_{\text{charb}}$ -implement our model using MSE loss instead of charbonnier loss, and 3) $w/o$ curriculum learning - training our model without curriculum learning. The 2$\times$ and 4$\times$ scale results on the IXI dataset are in Table \ref{table:2}. All three models perform worse than DisC-Diff, indicating that the components can enhance overall performance. The results of $w/o$ $\mathcal{L}_{\text{disent}}$ demonstrate that disentangling representations are effective in integrating multi-contrast information. $w/o$ $\mathcal{L}_{\text{charb}}$ performs the worst, consistent with our hypothesis that $\mathcal{L}_\text{MSE}$ is sensitive to noise in multi-contrast MRI and can cause the model to converge to local optima.
 \begin{table*}[h]
\centering
\caption{Ablation Study on the IXI dataset with $2\times$ and $4\times$ enlargement scale.}
\resizebox{0.7\columnwidth}{!}{%
\begin{tabular}{c||>{\centering\arraybackslash}p{1.2cm} >{\centering\arraybackslash} p{1.2cm}| >{\centering\arraybackslash} p{1.2cm} >{\centering\arraybackslash}p{1.2cm}}
\toprule
\hline
Scale            & \multicolumn{2}{c|}{2$\times$} & \multicolumn{2}{c}{4$\times$}    \\ \hline 
Metrics          & {PSNR} & {SSIM} & {PSNR} & SSIM \\ \hline \hline
$w/o$ $\mathcal{L}_{\text{disent}}$ & 37.15     & {0.9834}     & 31.08     & 0.9524 \\
$w/o$ $\mathcal{L}_{\text{charb}}$     & 36.70     & {0.9846}     & {31.05}     &    0.9532  \\
$w/o$ curriculum learning        & 37.58     & {0.9872}     &31.36     &  0.9533 \\ \hline \hline
\textbf{DisC-Diff (Ours)}  & \textbf{37.64}     & {\textbf{0.9873}}     & {\textbf{31.43}}     & \textbf{0.9551}   \\ \hline
\bottomrule
\end{tabular}
}
\label{table:2}
\end{table*}
 \section{Conclusion}
    We present DisC-Diff, a novel disentangled conditional diffusion model for robust multi-contrast MRI super-resolution. While the sampling nature of the diffusion model has the advantage of enabling uncertainty estimation, proper condition sampling is crucial to ensure model accuracy. Therefore, our method leverages a multi-conditional fusion strategy based on representation disentanglement, facilitating a precise and high-quality HR image sampling process. Also, we experimentally incorporate a Charbonnier loss to mitigate the challenge of MRI noise and outliers on model performance. Our extensive experiments highlight the potential of diffusion models as a new paradigm for multi-contrast SR and further clinical translation. Future work focuses on modeling diffusion processes of DisC-Diff on a compact and low-dimensional latent space instead of raw MRI images, potentially reducing memory consumption and training complexity.
%
%
%

\bibliographystyle{splncs04}
\bibliography{reference}
\end{document}